\documentclass{article}

\usepackage[preprint]{spconf}
\usepackage{hyperref}
\usepackage{amsmath,amssymb,amsfonts,mathrsfs,bm}
\usepackage{algorithmic}
\usepackage{graphicx}
\usepackage{textcomp}
\usepackage{xcolor}
\usepackage[numbers,sort&compress]{natbib}

\newcommand{\Fig}{Fig. }

\newcommand{\fourier}[1]{\mathscr{F}\left\{ #1 \right\}} 
\newcommand{\cosTrans}[1]{\mathscr{F}_c\left\{ #1 \right\}}

\newcommand{\Prob}{\mathbb{P}}
\newcommand{\distFunc}[2]{\text{dist}\left( #1, #2 \right)}
\newcommand{\enc}{\text{enc}}
\newcommand{\dec}{\text{dec}}
\newcommand{\wl}{\lambda}
\newcommand{\fl}{f}
\newcommand{\iu}{i}
\newcommand{\optFld}{U}
\newcommand{\objPln}{o}
\newcommand{\focPln}{f}
\newcommand{\arbFunc}{g}
\newcommand{\singCoord}{x}
\newcommand{\singFrCoord}{\omega}
\newcommand{\trCoord}{\mathbf{x}}
\newcommand{\frCoord}{\boldsymbol\omega}
\newcommand{\trCoordF}{\Tilde{\mathbf{x}}}
\newcommand{\refAmp}{A_r}
\newcommand{\ndims}{n}
\newcommand{\vecInd}{k}
\newcommand{\pol}{p}
\newcommand{\pixCoord}{\mathbf{u}}
\newcommand{\evThresh}{\delta}
\newcommand{\intens}{I}
\newcommand{\recIntens}{\Tilde{I}}
\newcommand{\logIntens}{L}
\newcommand{\policy}{\pi}
\newcommand{\sgn}{\text{sgn}}
\newcommand{\finalT}{T}
\newcommand{\evInd}{k}
\newcommand{\finEvInd}{K}
\newcommand{\dtInd}{s}
\newcommand{\finDT}{S}
\newcommand{\DTSet}{[\finDT]}
\newcommand{\nRows}{m}
\newcommand{\nCols}{n}
\newcommand{\imSet}{\mathbb{R}_{\geq 0}^{\nRows\times\nCols}}

\newcommand{\evCnt}{E}
\newcommand{\trunc}[1]{\left\lceil #1 \right\rfloor}
\newcommand{\recLogIntens}{\Tilde{\logIntens}}
\newcommand{\rowInd}{i}
\newcommand{\colInd}{j}
\newcommand{\dctIntens}{\mathcal{\intens}}
\newcommand{\dctRow}{q}
\newcommand{\dctCol}{r}
\newcommand{\dctSet}{\mathbb{R}^{\nRows\times\nCols}}
\newcommand{\rate}{R}
\newcommand{\dist}{D}
\newcommand{\simRate}{\rate_\text{sim}}
\newcommand{\simDist}{\dist_\text{sim}}

\title{RATE-DISTORTION ANALYSIS OF OPTICALLY PASSIVE VISION COMPRESSION}

\copyrightnotice{\parbox{\textwidth}{\copyright\ 2026 IEEE.  Personal use of this material is permitted.  Permission from IEEE must be obtained for all other uses, in any current or future media, including reprinting/republishing this material for advertising or promotional purposes, creating new collective works, for resale or redistribution to servers or lists, or reuse of any copyrighted component of this work in other works.}}

\twoauthors
  {Ronald Ogden, David Fridovich-Keil}
	{University of Texas at Austin\\
	Dept. Aerospace Engineering\\
	2617 Wichita St,\\
    Austin, TX 78712}
  {Takashi Tanaka}
	{Purdue University\\
	Aeronautics and Astronautics\\
    Electrical and Computer Engineering\\
	701 W Stadium Ave,\\
    West Lafayette, IN 47907}

\begin{document}
\ninept
\maketitle

\begin{abstract}
The use of remote vision sensors for autonomous decision-making poses the challenge of transmitting high-volume visual data over resource-constrained channels in real-time. In robotics and control applications, many systems can quickly destabilize, which can exacerbate the issue by necessitating higher sampling frequencies. This work proposes a novel sensing paradigm in which an event camera observes the optically generated cosine transform of a visual scene, enabling high-speed, computation-free video compression inspired by modern video codecs. In this study, we simulate this optically passive vision compression (OPVC) scheme and compare its rate-distortion performance to that of a standalone event camera (SAEC). We find that the rate-distortion performance of the OPVC scheme surpasses that of the SAEC and that this performance gap increases as the spatial resolution of the event camera increases.
\end{abstract}
\section{Introduction}
Vision sensors such as RGB cameras play a central role in robot perception and control. While vision provides a rich source of information for autonomous decision-making, transmitting this sensor data to remote users in real-time over resource-constrained communication channels poses challenges due to the high-volume and high-dimensional nature of vision data. This holds particularly in the context of robotics where strict latency requirements are often imposed, as in aerobatic maneuvering or high-speed collision avoidance \cite{mueggler2015towards,mueggler2014event}. Standard video codecs such as H.264 have been optimized to excel at efficiently compressing this dense data at the cost of computational complexity, usually involving intricate motion compensation computations as well as discrete cosine transforms (DCT) \cite{sayood2017introduction}. More recently, deep learning-based autoencoders have been proposed as an additional video compression solution \cite{habibian2019video,golinski2020feedback}. However, a forward pass through such networks involves many convolutional layers, resulting in nontrivial time complexity as well.

This work proposes a novel sensing paradigm in which an event camera observes the optically generated cosine transform of a visual scene to enable fast and efficient encoding. The proposed sensor performs purely hardware-level compression on the observed scene, eliminating the computational complexity associated with utilizing video codecs on raw frame-based footage. As with many video codecs, utilizing cosine transforms enables filtering of information based on spatial frequency. Event cameras encode changes in a scene rather than full frames of data, allowing them to operate at a significantly higher temporal resolution than frame-based cameras. This results in an efficient differential coding of the scene while maintaining a small sampling rate, which may ultimately enable closed-loop control of a high-speed control system. 

We develop a simple, noiseless event camera simulator that operates on grayscale frame-based videos in order to analyze the rate distortion performance of this novel sensor and compare it to that of a standalone event camera (SAEC). We find that our optically passive vision compression (OPVC) scheme outperforms an SAEC and that the performance gap increases with the spatial resolution of the event camera.

\section{Related Work}
Optical cosine transforms have been explored and implemented in multiple forms, often being applied for the purposes of image compression. \citet{gu1981optical} implemented an optical cosine transform by using a system of mirrors and prisms to evenly extend the input image and passing the result through a Hughes liquid crystal light valve before using a $4f$-system to obtain the corresponding Fourier transform. \citet{wong1992optical}  proposed using a microlens array and phase-conjugate mirror to enable an optical cosine transform aligned with the H.261 video codec. \citet{ma2019sar} utilized an optical cosine transform to aid in synthetic aperture radar image compression. These previous implementations of optical cosine transforms have only been utilized on static images, whereas  we seek to encode time-varying scenes with an event camera.

A rich ecosystem of literature on event cameras and their applications exists, much of which has been summarized in comprehensive surveys \cite{gallego2020event, chakravarthi2024recent}. To account for the lack of existing event camera datasets, many event camera simulators have been developed. \citet{bi2017pix2nvs} developed an algorithm to convert frame-based video to event streams with with multiple options for intensity comparison and timestamp determination. \citet{rebecq2018esim} developed an event camera simulator that adaptively samples photorealistic renderings to produce event streams, which \citet{gehrig2020video} used to produce event streams from frame-based videos. \citet{hu2021v2e} developed a toolbox to generate event streams from frame-based videos that accounted for realistic event camera phenomena including hot pixels, leak events and temporal noise. We implement a more simple, but highly vectorized simulator that easily integrates into our experimental framework.

\begin{figure*}[htbp]
\centerline{\includegraphics[width=\textwidth]{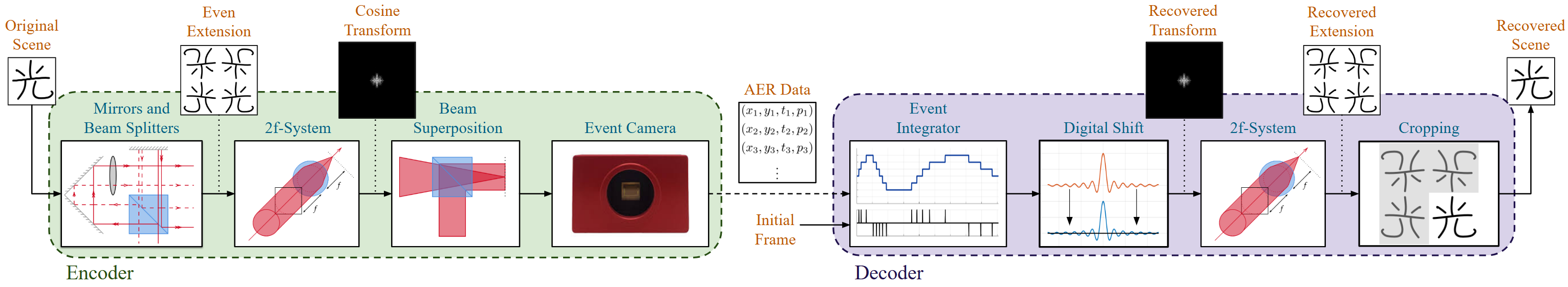}}
\caption{Block diagram of optically passive vision compression scheme.}
\label{fig:compOv}
\end{figure*}

\section{Compression Methodology}\label{sec:compMthd}
In this section, we provide background for the key elements of our proposed compression method. \Fig\ref{fig:compOv} summarizes the compression scheme. 

\subsection{Optical Cosine Transform}\label{sec:OCT}
Let the $\ndims$-dimensional Fourier transform of a function $g:\mathbb{R}^\ndims\to\mathbb{C}$ be defined as
\begin{equation}
    \fourier{\arbFunc}(\frCoord) := \int_{\mathbb{R}^\ndims} g(\trCoord)e^{-2\pi\iu\trCoord\cdot\frCoord}\,d\trCoord.
\end{equation}
Suppose one places a planar transparency a distance $\fl$ in front of a converging lens with focal length $\fl$ and illuminates it via a normally incident, coherent plane wave. If the transparency transmits the optical field $\optFld_\objPln(\trCoord)$, where $\trCoord\in\mathbb{R}^2$ is an element of the coordinate plane normal to the optical axis at the transparency, then the optical field in the back focal plane a distance $\fl$ behind the lens is given by
\begin{equation}
    \optFld_\focPln(\trCoordF) = \frac{1}{\iu\wl\fl}\fourier{\optFld_\objPln}\left(\frac{\trCoordF}{\wl\fl}\right), 
\end{equation}
where $\trCoordF\in\mathbb{R}^2$ represents a coordinate in the back focal plane and $\wl$ is the wavelength of the illuminating source \cite{goodman2005introduction}. While this enables one to optically compute the Fourier transform of an image, a camera only has the ability to sense information about the intensity of the optical field; one must also obtain the phase of the Fourier transform to recover the original image. 

One can easily show that for a real, even fuction $\arbFunc$,
\begin{equation}
    \fourier{\arbFunc}(\frCoord) = \int_{\mathbb{R}^\ndims} g(\trCoord)\prod_{\vecInd=1}^\ndims \cos(2\pi\singCoord_\vecInd\singFrCoord_\vecInd) \, d\trCoord := \cosTrans{\arbFunc}(\frCoord),
\end{equation}
where bold variables represent coordinates in $\mathbb{R}^\ndims$ and  $\singCoord_\vecInd$ and $\singFrCoord_\vecInd$ denote the $\vecInd^\text{th}$ elements of $\trCoord$ and $\frCoord$ respectively. We call $\cosTrans{\arbFunc}$ the cosine transform of $g$. Thus, the Fourier transform of a real, even function is real valued and equivalent to the cosine transform of that function. Any real function of compact support can be evenly extended such that it can be uniquely recovered from its even extension. Optically, one can do so using a system of mirrors and beam splitters \cite{gu1981optical}. 

The Fourier transform of the resulting even extension, being real, requires only binary phase recovery. Optically, one can achieve this by superimposing a plane wave $\optFld_r(\trCoordF) = \refAmp \in \mathbb{R}_{>0}$ onto the Fourier plane such that $\refAmp > \max_{\trCoordF} |\optFld_\focPln(\trCoordF)|^2$ as suggested in \cite{wong1992optical, ma2019sar}. Given an intensity measurement $I_m(\trCoordF) = |\optFld_\focPln(\trCoordF) + \refAmp|^2$, assuming $\refAmp$ is known, one can recover the transform using $\optFld_\focPln(\trCoordF) = \sqrt{\intens_m(\trCoordF)} - \refAmp$ since we select $\refAmp$ such that $\optFld_\focPln(\trCoordF) + \refAmp > 0$.

\subsection{Event Camera}
An event camera consists of an array of pixels that operate asynchronously. Each pixel only fires when the deviation since its last event in the log intensity of light it senses exceeds some threshold. That is, suppose a pixel's last event occurred at time $t_{\evInd-1}\in\mathbb{R}$ and that the log intensity of the light at that pixel over time is given by $L(t) := \log (I(t))$, where $I = |U|^2$ using our notation from Section \ref{sec:OCT}. Then the time of that pixel's next event is given by
\begin{equation}
    t_\evInd := \inf \{t : t>t_{\evInd-1}, |\logIntens(t)-\logIntens(t_{\evInd-1})|\geq\evThresh\},
\end{equation}
where $\evThresh$ represents a symmetric event threshold prescribed for the sensor. At time $t_\evInd$, this pixel sends the following address-event representation (AER) packet: $(\pixCoord_\evInd,\pol_\evInd,t_\evInd)$, where $\pixCoord_\evInd$ is the pixel position and $\pol_\evInd \in \{1, -1\}$ is the polarity of the change defined by
\begin{equation}
    \pol_\evInd := \sgn \left(\logIntens(t_\evInd) - \logIntens(t_{\evInd-1})\right).
\end{equation}

In the proposed compression scheme, the encoder transmits this stream of AER data to the decoder, where an event integrator reconstructs the scene recorded by the event camera with some loss. Given the sequence of event times $\{t_\evInd\}_{\evInd = 1}^\finEvInd$ and corresponding sequence of event polarities $\{\pol_\evInd\}_{\evInd = 1}^\finEvInd$ for a pixel whose initial sensed log intensity is known to be $L(0)$, the event stream can be integrated as follows to produce a lossy reconstruction of the pixel's log intensity history:
\begin{equation}
    \Tilde{\logIntens}(t) = L(0) + \sum_{\evInd:t_\evInd < t}\evThresh\pol_\evInd.
\end{equation}
In practice, such a naive reconstruction lacks robustness to sensor noise if full reference frames are not provided with sufficient frequency. However, we do not consider sensor noise, so this simple reconstruction method removes any confounding effects that a denoising filter could introduce to the study. Note that as one increases the event threshold $\evThresh$, the frequency with which events occur decreases, but the discrepancy between $L(t)$ and $\Tilde{L}(t)$ increases.

For the purposes of this work, we shall assume that each pixel's initial intensity is known to the decoder. Practically, this can be achieved by including a low-frame rate frame camera in the optical setup to provide periodic reference frames to the decoder, which could consequently add robustness to noise.

\section{Computational Analysis}
Our novel OPVC scheme and traditional video codecs are incommensurable; we have designed the former for computation-free compression of continuous scenes, and the latter was designed to optimize frame-based compression. In lieu of OPVC, one could observe a scene directly with a standalone event camera. In order to encode/decode a video with an SAEC, computational power is solely devoted to event integration. For the OPVC, optical components compute the cosine transform passively, leaving only a few trivial element-wise operations to be handled computationally during encoding/decoding in addition to event integration. Thus, per event, both compression schemes have the same order of computational complexity, and both perform hardware-level compression on continuous scenes, making them comparable. 

\subsection{Rate-Distortion Framework}
Event cameras transform a continuous visual scene into a discrete list of AER tuples, resulting in a distorted reconstruction. Such lossy compression methods lend themselves to comparison within a rate-distortion framework, which quantifies the trade-off between reconstruction quality (distortion) and the bits used over time to encode the video (rate) \cite{cover1999elements}. For a given event camera and temporal resolution, each AER tuple takes the same number of bits to store, so rather than bitrate, we use sampling rate for this analysis. While many distortion metrics fit this application, we use Multiscale Structural Similarity (MS-SSIM) \cite{wang2003multiscale}.

Formulating this idea precisely, let $\intens(t)\in\imSet$ denote the $\nRows$ by $\nCols$ image of pixel intensities at time $t\in\mathbb{R}$ of a visual scene. For this comparison, given a distribution $\Prob$ of scenes with time horizon $T$, we seek to compute a rate-distortion curve defined by
\begin{equation} \label{eq:contRD}
\begin{split}
    \rate(\dist; \policy) :=& \inf_\evThresh\mathbb{E}_{\intens \sim \Prob}\left[\frac{\finEvInd}{\nRows\nCols\finalT}\right], \\
    \text{s.t.} \quad &\mathbb{E}_{\intens \sim \Prob}\left[\frac{1}{\finalT}\int_0^\finalT\distFunc{\intens(t)}{\recIntens(t)}\,dt\right] \leq \dist, \\
    &\recIntens = \dec_{\policy(\evThresh)}(\enc_{\policy(\evThresh)}(\intens)),
\end{split}
\end{equation}
where $R$ is sampling rate, $D$ is distortion, $\policy$ represents the compression method (e.g. SAEC or OPVC) parameterized by event threshold $\evThresh$, $\enc_{\policy(\evThresh)}(\cdot)$ and $\dec_{\policy(\evThresh)}(\cdot)$ are the corresponding encoding and decoding functions respectively, $\finEvInd$ is the number of events that have occurred up to time $T$, $\text{dist}(\cdot)$ is our chosen distortion metric (MS-SSIM), and $\recIntens$ denotes the reconstructed scene.

\subsection{Simulation Methodology} \label{sec:simMthd}
In order to analyze the rate-distortion performance of both the SAEC and the OPVC, we implemented a simple event camera simulator. As inputs, the simulator takes an uncompressed, grayscale, frame-based video composed of frames $\intens (\dtInd)\in\imSet$, where $\dtInd\in\DTSet$ denotes the frame index, and an event threshold, $\evThresh\in\mathbb{R}_{>0}$. It returns an event count $\finEvInd$ corresponding to the total number of AER packets that the encoder sends to the decoder and a reconstructed video produced by the decoder, $\recIntens (\dtInd)$. While \eqref{eq:contRD} formulates the rate-distortion problem in continuous time, we shall adapt our treatment to fit the discrete nature of our simulation. 

Events in the simulator are counted as if the log-intensity of the continuous-time scene that the raw video frames represent varies monotonically and continuously between frames. We assume the temporal resolution and refractory period of each pixel are sufficiently small such that the pixel doesn't ``miss'' any events. These assumptions result in event counting similar to the event generation model in \cite{hu2021v2e}. The accuracy of this method improves when objects in the original frame-based video move slowly.

The simulation assumes the decoder has access to the initial frame $\intens(0)$ and accordingly sets $\recIntens(0) = \intens(0)$. The simulator converts frames to log-space by computing $\logIntens(\dtInd)$, the element-wise logarithm of $\intens(\dtInd)$. The decoder tracks a log-space reconstruction $\recLogIntens(\dtInd)$ as well. Suppose the decoder has reconstructed the frame $\recLogIntens(\dtInd-1)$. Using the assumption that a pixel's log-intensity varies monotonically between frames, all events that a particular pixel senses between frames will have the same polarity. Let $\evCnt:\DTSet\setminus \{0\}\to\mathbb{Z}^{\nRows\times\nCols}$ denote an event function representing the number of events that occur at each pixel after frame $s-1$ and up to frame $s$ multiplied by their respective polarities. This can be computed as follows:
\begin{equation}\label{eq:evFunc}
    \evCnt(\dtInd) = \trunc{\frac{\logIntens(\dtInd) - \recLogIntens(\dtInd-1)}{\delta}},
\end{equation}
where $\trunc{\cdot}$ denotes element-wise rounding toward zero. This matrix encodes all polarity and pixel location information for the events that occur between frames $s-1$ and $s$. Then, the decoder reconstructs $\recLogIntens(\dtInd)$ as $\recLogIntens(\dtInd) = \recLogIntens(\dtInd-1) + \evThresh\evCnt(\dtInd)$.

\begin{figure}[htbp]
\centerline{\includegraphics[width=0.48\textwidth]{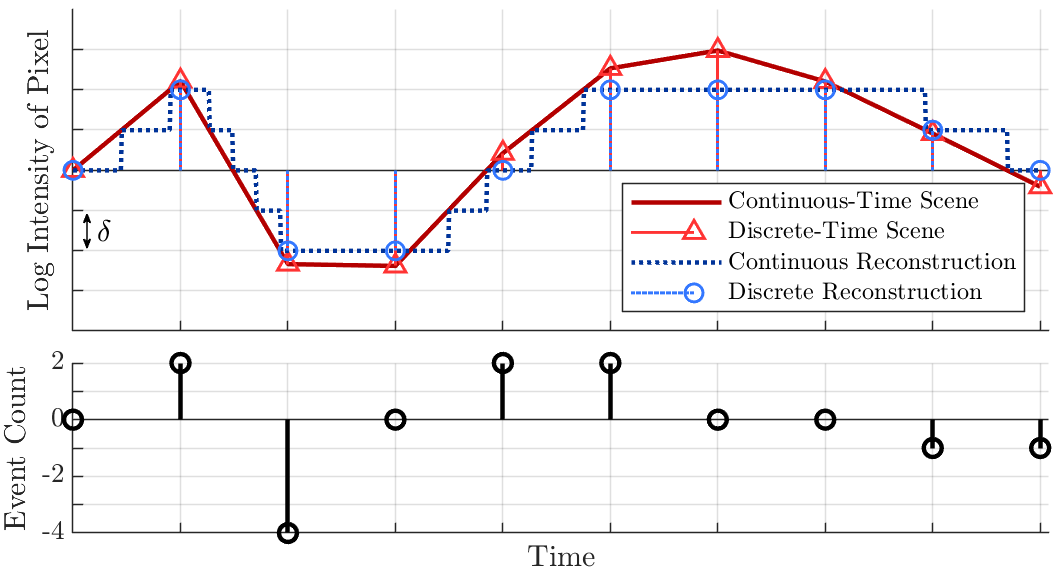}}
\caption{Illustration of the simulation methodology. The red triangles represent the log-intensity of the raw video data, $\logIntens(\dtInd)$. The solid red curve depicts the assumed continuous pixel variation between frames and the dotted blue curve represents the corresponding continuous reconstruction. The blue circles depict the frame-based reconstruction, $\recLogIntens(\dtInd)$. The lower plot depicts the event function $\evCnt(\dtInd)$ defined in \eqref{eq:evFunc}.}
\label{fig:simMethod}
\end{figure}

This simulation methodology is illustrated in \Fig\ref{fig:simMethod}. Letting $\evCnt_{\rowInd\colInd}(s)$ denote the element in row $\rowInd$ and column $\colInd$ of $E(s)$, the total number of events over the simulation is
\begin{equation}
    K = \sum_{i=0}^{m-1}\sum_{j=0}^{n-1}\sum_{s = 0}^{S-1} \left\vert \evCnt_{\rowInd\colInd}(s)\right\vert.
\end{equation}
As a final step, the simulator converts reconstructed log-intensity frames back into intensity frames. This simulator can analyze the performance of the SAEC, but the OPVC requires additional elements.

For the OPVC, the encoder utilizes a two-dimensional DCT to simulate the optical cosine transform. 
If $\dctIntens(\dtInd)\in\dctSet$ denotes cosine transform of $\intens(\dtInd)$ of the original video, then
\begin{equation}
    \dctIntens_{\dctRow\dctCol}(\dtInd) = \alpha_{\dctRow} \alpha_{\dctCol} \sum_{\rowInd = 0}^{\nRows-1}\sum_{\colInd = 0}^{\nCols-1} \intens_{\rowInd\colInd}(\dtInd) \cos \frac{\pi(2\rowInd+1)\dctRow}{2\nRows} \cos \frac{\pi(2\colInd+1)\dctCol}{2\nCols},
\end{equation}
where $\alpha_\dctRow = \sqrt{1/\nRows}$ for $\dctRow = 0$, $\alpha_\dctRow = \sqrt{2/\nRows}$ for $\dctRow>0$ and $\alpha_\dctCol$ is defined analogously. The OPVC simulator shifts this transformed video by a positive constant large enough that the result is positive for every pixel over all frames before passing it to the simulated event camera. The decoder subtracts the same constant from the reconstructed output of the event camera. These constant shifts emulate the plane wave superposition for phase recovery discussed in Section \ref{sec:OCT}. Finally, we compute the inverse DCT of the result to obtain the final reconstruction, $\recIntens(\dtInd)$.

Given a single video, the simulation computes the following sampling rate and distortion:
\begin{align}
    \simRate(\intens; \policy(\evThresh)) &= \finEvInd/(\nRows\nCols\finDT), \\
    \simDist(\intens; \policy(\evThresh)) &= \frac{1}{\finDT}\sum_{s = 0}^{S-1} \text{MS-SSIM}\left(\intens(\dtInd),\recIntens(\dtInd)\right).
\end{align}
For each compression method, we computed the average of $\simRate$ and $\simDist$ over a collection of videos. These computations were repeated over a range of event thresholds to construct a rate-distortion curve. We performed this analysis at varying video resolutions by spatially downsampling the dataset to the desired resolution via block-by-block pixel averaging. We used the Ultra Video Group (UVG) dataset for analysis \cite{mercat2020uvg}. This dataset, developed for the intent of video codec analysis, consists of 16 uncompressed videos at a resolution of $3840 \times 2160$ with up to 600 frames. 

\subsection{Results}
\begin{figure}[htbp]
\centerline{\includegraphics[width=0.48\textwidth]{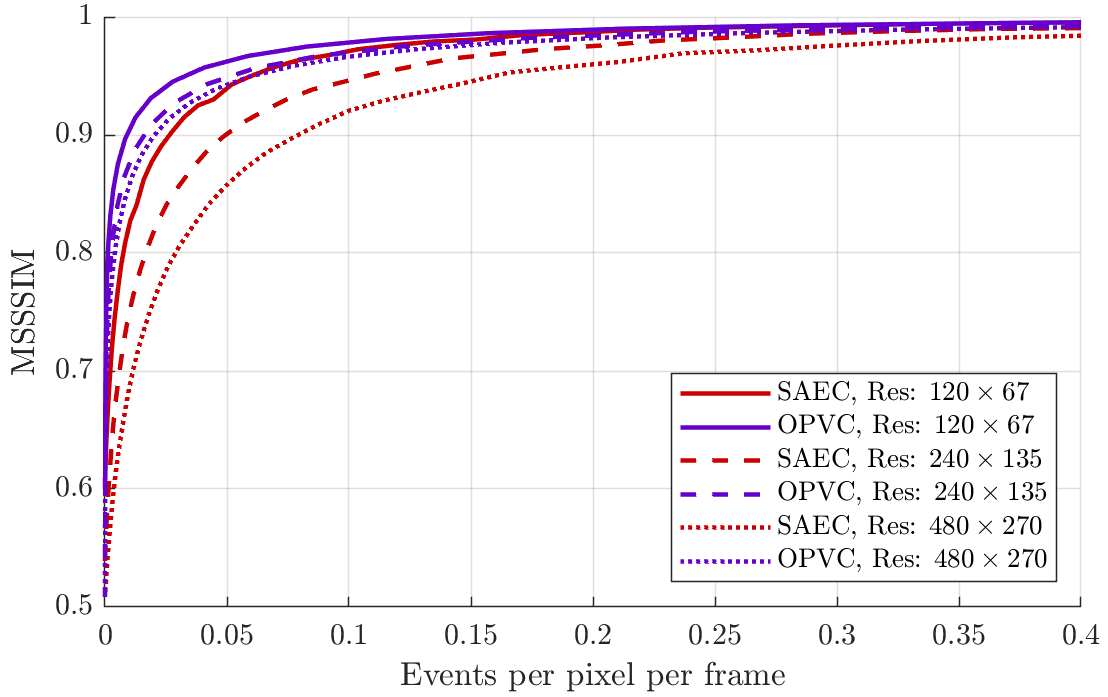}}
\caption{Rate-distortion performance of SAEC and OPVC on UVG dataset over multiple spatial resolutions.}
\label{fig:resComp}
\end{figure}

\Fig\ref{fig:resComp} displays the results of the simulation described in Section \ref{sec:simMthd}. At each resolution tested, the OPVC achieves a lower sampling rate per pixel than the SAEC at any given distortion constraint. Moreover, the per-pixel performance gap between the OPVC and SAEC increases with the as the resolution of the event camera increases. Spatial downsampling inherently filters out high spatial frequency information from the observed scene, which can minimize the benefits of the OPVC over the SAEC at lower resolutions. We performed similar experiments with varying distortion metrics such as mean-squared error, SSIM and peak signal-to-noise ratio, all of which showed the same qualitative results for this data set.

\begin{figure}[htbp]
\centerline{\includegraphics[width=0.4\textwidth]{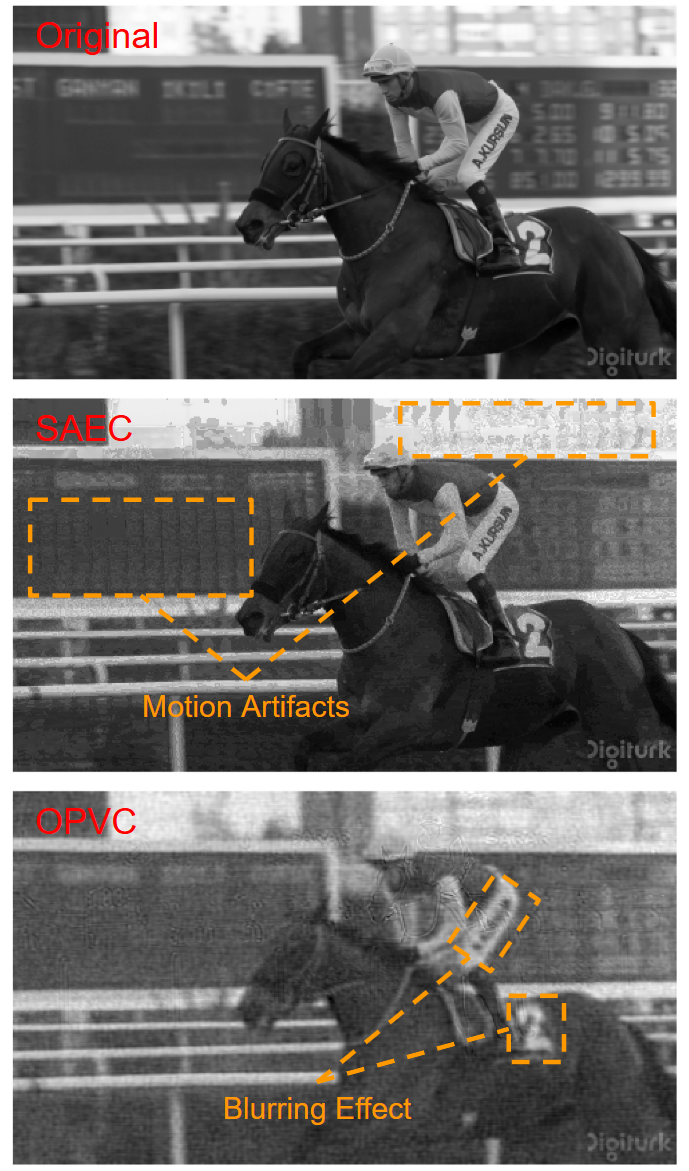}}
\caption{Qualitative comparison of compression results. The top image is a frame from the original video, which depicts a jockey riding a horse to the left as the camera pans with him. The middle image is a frame reconstructed using an SAEC at 0.9 MS-SSIM. The bottom image is a frame reconstructed using an OPVC at 0.9 MS-SSIM.}
\label{fig:visComp}
\end{figure}

\Fig\ref{fig:visComp} shows a qualitative comparison of the SAEC and OPVC simulations. While the SAEC appears to retain many sharp edges of the original scene, it contains distortion artifacts due to the motion of edges between objects of high contrast. The OPVC reconstruction does not appear to have these motion artifacts, but does appear to have high frequency noise, blurring sharp edges that existed in the original scene. This occurs because the optical cosine transform concentrates most of the energy of the image toward the low frequency domain, causing the event camera to filter out high-frequency information.

\section{Conclusion}
In this work, we proposed a high-speed, computationally minimal compression scheme that combines Fourier optics with event cameras to passively filter high spatial frequency information from a visual scene. Our simulation of the proposed compression method shows that it outperforms standalone event cameras in a rate-distortion sense. While these initial results show promise, they represent an ideal scenario. Future studies should include event camera sensor noise, higher fidelity conversions of frame based videos to event data (see \cite{bi2017pix2nvs}, \cite{hu2021v2e} and \cite{zhang2024v2ce}) and account for the rate-distortion tradeoff that occurs with the addition of periodic initializing frames.

\bibliographystyle{plainnat}
\bibliography{references}

\end{document}